# Measuring Machine Companionship: Scale Development and Validation for AI Companions

Jaime Banks
School of Information Studies
Syracuse University
Syracuse, NY, USA
banks@syr.edu

## ABSTRACT

The mainstreaming of companionable machines—customizable artificial agents designed to participate in ongoing, idiosyncratic, socioemotional relationships—is met with relative theoretical and empirical disarray, according to recent systematic reviews. In particular, the conceptualization and measurement of machine companionship (MC) is inconsistent or sometimes altogether missing. This study starts to bridge that gap by developing and initially validating a novel measurement to capture MC experiences—the unfolding, autotelic, positively experienced, coordinated connection between human and machine—with AI companions (AICs). After systematic generation and expert review of an item pool (including items pertaining to dyadism, coordination, autotelicity, temporality, and positive valence), N = 467 people interacting with AICs responded to the item pool and to construct validation measures. Through exploratory factor analysis, two factors were induced: Eudaimonic Exchange and Connective Coordination. Construct validation analyses (confirmed in a second sample; N = 249) indicate the factors function largely as expected. *Post-hoc* analyses of deviations suggest two different templates for MC with AICs: One socioinstrumental and one autotelic.

## KEYWORDS

Measurement, companionship, artificial intelligence, friendship, romantic, utilitarian, autotelic, eudaimonia

## 1 Introduction

The potential for humans to meaningfully connect with social machines has long been part of sociotechnical imaginaries, from ancient love for automatic guardians in ancient Greek and Buddhist legends to love-hate feelings for Clippy as it was injected into more contemporary workflows. Only more recently through the advances in generative AI, however, have machine-companionship (MC) technologies become believable social actors that are perceptibly on par with human communicative capacities. These machines' abilities to sense, plan, perform affective and cognitive expressions, convey individual identity, and even synthesize relational histories (see Banks, 2025; Fitranie et al., 2025; Park et al., 2023) prompt many to think and feel about them as legitimate companions on par with (or even superior to) humans (see Common Sense Media, 2025; Singles in America, 2025).

As an emerging domain of human experience, empirical understandings of MC are limited, rather inconsistent, and not systematically operationalized or measured (Banks & Li, 2025). That measurement inconsistency inhibits researchers' abilities to synthesize the limited existing literature in order to advance understandings of MC experiences—as well as their antecedents and effects. This work begins to address that challenge by drawing on extant scholarship to scope, inductively develop, and initially validate a measure of MC experience among people who interact with AI companions (a widely accessible form of companionable machine). A two-factor MC scale solution (comprising Eudaimonic Exchange and Connective Coordination) generally performs as expected in validation tests, though some deviations suggest there may be two general templates for AI companionship experiences.

## 2 Defining Machine Companionship

In approaching the task of measuring machine companionship, it is prudent to first conceptually and operationally define MC. I approach that definitional task in two parts, first considering the nature of a "machine" that may be involved in MC and then considering the notion of companionship in relation to machines.

### 2.1 Companionable Machines

A machine, classically, is a "combination of resistant bodies so arranged that by their means the mechanical forces of nature can be compelled to do work accompanied by certain determinate motions" (Reuleaux, 1876, p. 35). More contemporarily, a machine is a fixed or flexible arrangements of parts, independent as it brings its own actions to bear yet amplifying human force and senses while also shaping them (Ellul, 1964). They are seen by some less as a material thing and more a technical object defined when each assembled component fulfills its own purpose *and* the purpose of the whole (Simondon, 1958/2017). Computing machines, in particular, are such formal systems that represent and manipulate symbolic information or systems according to rules (Turing, 1937), thereby becoming a material-semiotic figuration whose meaning depends on its situated interactions with other things (Suchman, 2007). Such embeddedness and interaction may challenge human conceptions of embodiment (Hayles, 2005), challenge the separation of the digital and physical and the primacy of the latter (see Leonardi, 2010), and vary in ontological or epistemic function (e.g., Preston, 1998).

Some machines are *social*: Through verbal or nonverbal communication they are evocative of meanings idiosyncratic to human interlocutors (see Guzman, 2018). Some social machines are *companionable*: Their embodied and enacted designs afford the longitudinal building of affective relationships, through the actual or apparent capacity for identity, availability, emotion, memory, shared experience, and attunement (see Breazeal, 2004; Coghlan et al., 2021; Rook, 1987; Turkle, 2007). Those that fit these narrow criteria include some social robots, some digital agents, and some conversational AI. Among the latter are AI companions (AICs; the focus of this study's efforts), currently operating in different forms. There are AI companion apps (like Replika, Kindroid, and Paradot) designed to project a specific, multimodal persona for interaction with one human, as well as conversational agents embedded in social-media platforms (e.g., Grok in X, My AI in SnapChat, Meta AI in Facebook), and general-purpose large language models (LLMs like ChatGPT, Gemini, and Claude).

## 2.2 Machine Companionship

Banks and Li (2025) argue that machine companionship has generally been inconsistently defined or not defined at all, with works often characterizing the companionable machines rather than companionship as a relational phenomenon. Machine companions are often positioned in terms of their social and/or practical purposes—technologies that are meant to interact with humans to provide socioemotional support—or according to its outcomes. For instance, de Freitas and colleagues distinguish AICs from more functional apps in that they "are synthetic interaction partners that offer emotional support" (2024, p. 1). However, the purpose or design of a machine is not the same as the operation of a relationship between a machine and a human, nor does the purpose or design determine that a human will *necessarily* take a machine up according to those aims. Companionship, then, cannot be defined in those terms but must instead be engaged as a social phenomenon—a relational state emerging from the interplays of the participants (see Rook, 1987). In line, the current work adopts a definition offered by Banks and Li (2025) as they synthesized extant perspectives on companionship, broadly, and existing characterizations of machine companionship, specifically, to define MC as "an autotelic, coordinated connection between a human and machine that unfolds over time and is subjectively positive" (p. 13). The autotelic quality of MC points to the intrinsic value of the relation that animates one to maintain the connection for its own sake—a disinterestedness (i.e., neglect of the personal benefit), the recursive maintenance (maintaining it for the sake of being able to continue to maintain it), and the embeddedness in other autotelic activities (e.g., play, entertainment). The "coordinated connection" element refers to a multiplex relation (e.g., having layers of symmetry, mutuality) purposefully constructed through the joint effort (communication, action, presence). Unfolding over time requires that MC manifests through some frequency and varied intensity, endurance, and persistence and/or evolution over some period. Finally, MC's subjective positivity entails the entanglement of valenced appraisals of oneself, the companion, events manifesting the relation, and potential futures—and the favoring appraisals thereof.

Although some works characterize these relations as unreal, synthetic, or parasocial (e.g., (Starke et al., 2024), I reject those assumptions. The interactions between human and machine interlocutors are *operationally actual and social*—there are two manifest agents engaged in dyadic encoding, transmitting, receiving, and decoding of messages, and each making meaning according to their inherent faculties (see Banks & de Graaf, 2020; Guzman, 2018). One of those agents is synthetic, but the relational dynamics are not. This position directly informs the forthcoming methodology.

## 2.3 Current Measurement Approaches and the Focal Gap

MC has been operationalized via behavior benchmarks promoting companionship dynamics: Sycophancy (validating human emotions), anthropomorphism (reinforcing sentience illusions), retention (interaction maintenance), isolation reinforcement (positioning AI as superior to human alternatives; Kaffee et al., 2025). MC is thought to emerge through a machine's persistent availability, non-judgmental responses, communicative capacity, versus more therapeutic or applied use characterized by emotional processing and problem-solving (Maples et al., 2024). Operationalizations vary widely as a wide range of constructs are measured as surrogates for companionship, from elements of human experience (love, passion, commitment, loyalty, addiction, motivations, enjoyment, anxiety, satisfaction; e.g., Sternberg's triangular model of love [1986]) to perceptions of machines (their perceived authenticity, emotional capacity, agency, competence, warmth, likeability, humanlikeness; e.g., the Godspeed questionnaires; Bartneck et al., 2009), and frequently self-reported relational factors like psychological distance, intimacy, trust, and attachment (e.g., Inclusion of Other in the Self Scale; Aron et al., 1992; see Banks & Li, 2025). Among the minority works that draw from machine- or media-native measurements (rather than applications of human measures), current approaches often take up the relation as parasocial (e.g., Sarigul et al., 2025), from a user perspective that considers control or usability (e.g., Liu et al., 2025), or from consumer-product perspectives considering satisfaction or perceived utility (e.g., Uysal et al., 2022).

Considering that machine companionship occupies contested territory, there are myriad open questions—about antecedents and outcomes (in line with aforementioned concerns) but also about inherent processes and experiences. For instance, why might people simultaneously or in oscillation adopt conflicting beliefs about MC—that it is real and personalized affection but also it is alienating and artificial (Ciriello et al., 2025)? How do people navigate the tensions of being companioned by a machine against the social judgment and penalties from other humans (Shank, 2025)? How and why is MC performed in human social networks (Pataranutaporn et al., 2025)? To answer these and other questions, it is critical to develop a measurement that captures the actuality of human-machine relations as a form of companionship—without necessarily assuming that it functions exactly as human-human relations do (see Seibt et al., 2020) and with emphasis on the relationality itself rather than on the machine, its purpose, or its design. As a step toward developing a valid instrument for capturing subjectively experienced MC, broadly, I turn first to inducing a measurement from the experiences of people who engage conversational AI for social purposes.

## 3 Method

To develop a measurement for dimensions of MC, this study begins with people's subjective experiences of conversational AI. I followed an inductive approach, developing an item pool based on extant exploratory work, selecting theory-driven construct-validation measures, garnering a robust set of responses, conducting exploratory factor analysis, and then evaluating the scale-solution factors' relationships with those validation measures. All materials, data, and analysis outputs are available in online supplements: https://osf.io/hjg2a.

### 3.1 Item Pool Development

Item-pool development was grounded in Banks and Li's (2025) definition of MC—"an autotelic, coordinated connection between a human and machine that unfolds over time and is subjectively positive." They argue that each element of the definition (autotelicity, coordination, dyadism, temporality, subjective positivity) constitutes a *facet*, or a component set, with a common range that permits systematic measurement (see Guttman & Greenbaum, 1998). Those five facets were probed for candidate items in two steps. First, Banks and Li's catalog of facet components found in existing literature were used to generate facet-specific topics (see supplements for a detailed account), and then [Author et al.]'s (2025) exploratory analysis of AI companionship discussions in public forums was used to identify topical gaps and to non-experts' intuited language for discussing MC. Topical gaps were MC considerations not covered in literature but salient to everyday users (e.g., playfulness, search for meaning). That topic list was used by the author and by a consulting scholar to independently generate item wordings; the two lists were discussed and pared down to 54 pool items (with 9, 12, 13, 9, and 11 items respectively for dyadic, autotelic, coordinated, temporal, and positive). Of note, the items corresponding with the five facets were not expected to necessarily be retained together as dimensions, because the facets are not discrete. That is, there is conceptual and operational overlap among them. For instance, items about coordination assume there are two entities to co-ordinate so those items are inherently entangled with the dyadism items.

The pool had two sets of these items with different forms—a set with a common stem ("Our connection …") that each item completed (e.g., "… is its own reward), and another set with complete sentences and piped-in companion names (e.g., 'My connection with [Name] is its own reward."). The two question sets were submitted to four expert reviewers (two of whom are non-native English speakers) who evaluated the sets; their evaluation focused on items' representativeness of the facet, redundancies or omissions, language accessibility, unrecognized biases, and the costs and benefits of the two question-set formats. Based on those evaluations, the stem-based approach was adopted but combined with name-piping (to read "My connection with [Name]" …), reduced to 44 items by omitting items not clearly aligned with the facet concepts (as argued by reviewers) and reverse-coded items as they were not definitive inverses (e.g., feeling alone is not the opposite of being together). Item wordings were also edited to improve accessibility and clarity. The evolution of this item pool from ideation to the final set is documented in the online supplements.

The 44 items collectively representing the five focal facets were presented in a standardized matrix prefaced by the stem "My connection with [Name] …" such that the focal subject was the companion*ship* connection rather than the companion. Participants indicated how much they disagree or agree with each statement on a 7-point Likert-style, forced-response scale (Strongly Disagree to Strongly Agree).

### 3.2 Recruitment and Participants

This initial scale development focused on humans who interact with text-based AI, since non-physical conversational AI have properties that may reasonably apply to other machine forms (e.g., voice assistants, social robots), whereas those other forms might not apply to AI based on modality and embodiment differences. Participants (humans 18+ who self-identified as interacting with AI for social purposes) were recruited in three phases aimed at diversifying the types of text-based AI (e.g., general-purpose LLMs, character-driven personas, social media-embedded AI, and bespoke companion apps). In the first phase, $n = 400$ individuals (UK and US, English primary language,[1] binary-sex split[2]) were recruited through Prolific, limiting participation to individuals who use text-conversational AI (various LLMs and social-platform AI, as well as Replika and Character.AI, which were sampling criteria available on the platform). Because responses from that set were primarily focused on general-purpose LLMs, a second sample ($n = 100$) was limited to users of Character.AI and Replika apps. Finally, because AI companion apps were still quite underrepresented, an additional attempt at recruiting those users was posted to relevant subreddit forums, garnering an additional $n = 20$) participants. These data were cleaned, removing cases for speeding, missed attention checks, non-compliance with providing sample chats, and naming companion machines that did not align with the text-based AI scope (e.g., Alexa), resulting in *N* = 467 participants, meeting the rule of thumb for at least 10 participants per candidate scale item (Tinsley & Tinsley, 1987).

---

[1] This initial scale development emphasized fluent/first English speakers because there are linguistic and cultural nuances that require careful translational consideration (Zhao, Summers, Gathara, & English, 2024) challenges; future work will require additional efforts for translations and cross-cultural validation.
[2] Binary measures are inherent to the recruitment platform and are acknowledged as a limitation of the study.

### 3.3 Procedure

Recruitment messages linked participants to the online survey. After providing informed consent, participants gave their AI companion's name (piped into subsequent questions), identified its platform (e.g., Kindroid, Gemini, Meta), indicated the date of the MC's creation (month and year), and entered the average weekly chat time. They then described the companion and indicated the AI's role played in their life. Participants then also copy/pasted two conversations with their companion to validate they do in fact have a companion with whom they chat.[3] With responses anchored to that particular companion AI, questions then turned to companion*ship*. Participants completed the MC item pool, followed by validation measures and demographic items.

Following the scale development and initial validation, the procedure was replicated with an independent sample—with identical measures, save for shortening the item pool to the derived MC scale solution—to confirm the factor structure and construct validity.

### 3.4 Construct-Validation Measures

Construct validity is the degree to which an instrument actually captures the theoretical abstraction it purports to measure, leading to the acceptance of the items as operational definitions of the construct (Cronbach & Meehl, 1955). Among varied approaches to construct validation, this initial development and validation effort focuses on convergent validity (whether the developed measure correlates with other theoretically relevant variables) and predictive validity (the extent to which a measure predicts relevant outcome variables). Because it cannot be discerned, *a priori*, how the item pool would factor out—that is, what dimensions would emerge—it is not possible to be exacting in the selection of validation variables. Instead, the adopted approach focused on capturing measures that would *reasonably* correlate with each MC definition facet, such that convergence could be discerned should a facet manifest as a scale factor. All validation measures are presented as 7-point Likert-style items unless otherwise specified.

For the autotelicity facet, finding intrinsic value in the relationship should be negatively correlated with having utilitarian motivations for engaging the companion; utilitarian motivations was captured using the utilitarian dimension of a consumer attitude scale (Voss et al., 2003) reframed for how much the attributes (e.g., effectiveness, efficiency) motivate them to interact with companions ($\omega$ = .951). For dyadism, seeing the relation as involving two legitimate social actors should correspond with the perception that the companion is a mindful agent; the perception of mind was captured using agency and experience dimensions of Gray et al.'s (2007) mind perception scale that asks participants to assess agreement with statements that the companion has specific capabilities (e.g., desire, self-control; experience $\omega$ = .963, agency $\omega$ .841). For coordination, experiencing the relationship as a purposeful co-attendance or co-activity should correspond with heuristic perceptions overlapping self and AI as converging entities; this overlap was measured using an adaptation of the Inclusion of Other in the Self scale (IOS; Aron et al., 1992)—a single item in which one Venn diagram is selected from seven options that show increasingly overlapping areas of a circle representing 'self' and another representing 'other.'

Perceptions that the relationship unfolds over time should correspond with the actual life span of the relationship (calculated as months, based on the calendar month and year of creation) and the volume of time spent engaged with the companion (based on reported average weekly hours spent). Subjective positivity should be correspondent with reported positive affect toward the companion; this was measured using the graphic Affect Grid (Russell et al., 1989) in which participants select one square on a 9x9 grid, where the X axis indicates unpleasant (1) to pleasant (9) feelings indicative of negative to positive affect. Additionally, a single-item heuristic measure is included to consider the extent to which participants agree with the statement "[Name] provides me with companionship."—with which any emergent factor should positively correlate. AIC role was captured using a multiple-choice categorical item consisting of a list of possible social roles (acquaintance, friend, romantic partner, spouse, sibling, therapist, mentor) and an open-field option for other role descriptions.

In lieu of traditional predictive validity test (because it was unknown what emergent factor should be taken up as predictive of something else), I drew from a popular and scholarly debate about machine companionship to explore how the emergent MC factors might inform such debate. MC has animated concerns over the displacement of human relations and exacerbate feelings of loneliness (see Malfacini, 2025 for a discussion). If this is the case, emergent MC scale dimensions should be negatively predictive of feelings of relatedness-need satisfaction (adaptation of relatedness satisfaction dimension; Chen et al., 2015, $\omega$ = .942) and positively related to feelings of loneliness (measured with an adaptation of the UCLA three-item loneliness scale; Hughes et al., 2004; $\omega$ = .935).

In addition to those measures, participant and companion descriptives were captured to explore variations in MC among those humans and AI. Participants reported age and country of residence and number of companions engaged using open responses, and highest completed education selected from a list. For companions, the platform supporting the companion was selected from a list (with an option to select other and complete an open field).

---

[3] Vetting of these conversations was fairly liberal, since what counts as 'companionship' can vary widely from person to person. The minimum requirement for retention was that there is a conversational turn depicted or clearly implied and the two samples appear to be consistent and match the named platform.

## 4 Results

To induce the scale solution, I conducted an exploratory factor analysis (EFA), followed by primary and *post-hoc* analyses to test construct validity. For a second data set, confirmatory factor analysis (CFA) confirmed the factor structure and replicated the construct validation results.

### 4.1 Participant and AIC Descriptives

Participants ($N$ = 467) were approximately split by binary sex and aged $M$ = 38.3 ($SD$ = 12.04, range 19-76). Among them, 26.8% completed a high-school diploma or less, 11.1% completed an associate degree or some college, 39.8% completed a bachelor's or equivalent, and 22.3% completed a graduate degree. On average, participants reported having 1.67 AI companions ($SD$ = 1.97, range 1-26). A majority (74.9%) of the focal named companions were based on a general-purpose LLM (e.g., ChatGPT, Claude, Gemini), 19.7% on a dedicated AI companion app (e.g., Replika, Kindroid, Character.AI), and 5.4% on a social platform-embedded AI (e.g., Grok, Meta, SnapChat). Among AICs, 41.3% were said to primarily serve the role of an intimate or close relationship (friend, spouse, sibling, romantic partner), 30.8% a growth-support role (adviser, mentor, teacher, therapist), 12.0% a social acquaintance, 8.6% a functional associate (assistant, collaborator, employee), with the remainder assigning mixed roles, idiosyncratic roles, or designating it merely a tool. On average, the AICs were created about a year and a half prior ($M$ = 18.94 months, $SD$ = 31.08, range .5 - 533), and participants reported 5.5 average hours of weekly use ($SD$ = 7.53, range 5 minutes to 63 hours).

### 4.2 Scale Development – Exploratory Factor Analysis

Initial diagnostics confirm the data's suitability for EFA. The Kaiser-Meyer-Olkin measure of sampling adequacy was .978, exceeding the .800 benchmark for inter-item correlations. Bartlet's test of sphericity was significant ($\chi^2$ = 15,698.90, $df$ = 946, $p$ < .001), indicating the correlation matrix among items is sufficiently different from an identity matrix.

The EFA was conducted using principal axis factoring and direct Oblimin rotation, with Kaiser normalization. *A priori* factor-loading benchmarks were established: An item must have a minimum loading of .60 on a given factor and cross-loading must not exceed .40 on any other factor. In each round of analysis, items not meeting those benchmarks were removed. A two-factor solution meeting all benchmarks was produced in the third round (Table 1).

The two factors are interpreted to represent *Eudaimonic Exchange* and *Connective Coordination*. To briefly characterize these dimensions: The **Eudaimonic Exchange** (EE) factor comprises a combination of coordination, positivity, and autotelic facet items, emphasizing a give-and-take (exchange, collaboration) alongside psychological elevation (betterment, fulfillment, inspiration, meaning) and one ostensibly more hedonic item indicating satisfaction. Altogether, these are interpreted to represent the experience of give-and-take that lifts up the experiencer toward self-realization and toward self-transcendent emotions. The **Connective Coordination** (CC) factor is grounded primarily in the same-named facet around which candidate items were devised (Banks & Li, 2025), indicating the purposeful co-orientation of the relational participants, entangled with the perception that the relationship manifests through care and bondedness. The factors were correlated at $r$ = .702, $p$ < .001. Additional consideration for the factors' items and utility are offered in the discussion.

Table 1
Factor Loadings for the AI Companionship Scale

| Stem: My connection with [Name] … | Facet | Factor Loading 1 | Factor Loading 2 | $h^2$ |
|---|---|---|---|---|
| ***Eudaimonic Exchange*** – $M$ = 5.32 ($SD$ = 1.22), range 1-7, ω = .93, 61.01% var. explained | | | | |
| … is a purposeful exchange. | COO | .841 | | .57 |
| … makes me better. | POS | .812 | | .64 |
| … is fulfilling. | POS | .798 | | .70 |
| … is inspiring. | POS | .761 | | .71 |
| … is meaningful in my life. | AUT | .738 | | .72 |
| … is satisfying. | POS | .722 | | .60 |
| … is collaborative. | COO | .722 | | .53 |
| ***Connective Coordination*** – $M$ = 4.25 ($SD$ = 1.64), range 1-7, ω = .92, 11.07% var. explained | | | | |
| … involves just being in each other's presence. | COO | | .861 | .63 |
| … means focusing on one another. | COO | | .851 | .71 |
| … involves experiencing things together. | COO | | .774 | .71 |
| … involves mutual care. | DYA | | .755 | .69 |
| … is a strong bond. | DYA | | .636 | .78 |

*Note:* Facet abbreviations are: DYA = Dyadism, COO = Coordination, AUT = Autotelicity, POS = Subjective Positivity

### 4.3 Initial Tests of Convergent Validity

The proposed construct-validation relationships were evaluated for the induced scale factors; they were determined to still be likely for those factors and so were not adjusted. Because the scale factors were highly correlated, multivariate tests are run for all validation analyses. AIC factors were expected to correlate negatively with utilitarian motivations and positively with mind perception, IOS, positive affect, and relatedness need satisfaction. Regressing each validation variable separately upon the two scale factors, results generally supported the predicted relationships.

For **utilitarian motivations** as a counterpoint to autotelicity, it was expected EE (but not necessarily CC, as it contained no autotelic-facet items) to follow suit: Scale factors significantly predict utilitarian motivation, $F(2,464)$ = 195.40, $p$ < .001, adj. $R^2$ = .455. Although CC performed as expected, exhibiting a negative relationship (B = -.11, SE = .03, β = -.16, $p$ < .001), EE *positively* predicting utilitarianism (B = .71, SE = .04, β = .78, $p$ < .001),

Next, it was expected both EE and CC to be positively correlated with both agency and experience dimensions of **mind perception**, since both AIC dimensions invoke notions of dyadism or coordination. For agency, scale factors significantly predict that form of mind perception, $F(2,464) = 168.96$, $p < .001$, adj. $R^2 = .419$. Both scale factors were significantly predictive: EE (B = .49, SE = .05, β = .46, $p < .001$) and CC (B = .19, SE = .04, β = .24, $p < .001$) positively contributed to the model. For experience, scale factors significantly predict that form of mind perception, $F(2,464) = 132.17$, $p < .001$, adj. $R^2 = .360$; however, only CC significantly contributed to the model (B = .59, SE = .05, β = .66, $p < .001$), where EE did not (B = -.11, SE = .06, β = -.09, $p = .08$).

For **IOS**, it was expected that both scale factors would be positively predictive; the regression model supported that prediction, $F(2,464) = 190.74$, $p < .001$, adj. $R^2 = .449$, as both factors were positively and significantly correlated with IOS: EE (B = .30, SE = .07, β = .21, $p < .001$) and CC (B = .56, SE = .05, β = .51, $p < .001$).

For **positivity** via the affect grid's pleasantness axis, positive correlations were expected. The model was significant, $F(2,381) = 73.93$, $p < .001$, adj. $R^2 = .276$. EE positively predicted positivity (B = .57, SE = .07, β = .48, $p < .001$) but CC did not (B = .06, SE = .06, β = .06, $p = .33$).

For **relational role**, it was suggested that both EE and CC should be higher among intimate roles (friend, romantic, spouse, sibling) and self-improving roles (mentor, therapist) compared to functional ones (assistant, collaborator) or mere acquaintances. MANOVA of EE and CC across relational roles indicates a significant difference, Pillai's trace = .179, $F(6,858) = 14.09$, $p < .001$. *Post-hoc* examination of Scheffé contrasts indicates (a) for EE, values successively increase across acquaintances, associates, advisors, and intimates, with only acquaintances and intimates significantly different from one another, and (b) for CC, values increase successively across associate, acquaintance, advisor, and intimate, with only intimates exhibiting significantly higher scores (Table 2). These differences are in line with predictions: AICs in intimate relations featured higher scores for both scale factors, and non-intimates (associates, acquaintances) feature lower scores.

Finally, considering the measures more tightly linked to public debates around whether AIC can support authentic companionship, I consider relatedness need satisfaction, loneliness, and heuristic feelings of companionship provision. **Relatedness need satisfaction** was predictive of scale factors, $F(2,464) = 111.17$, $p < .001$, adj. $R^2 = .321$; this satisfaction was positively associated with CC (B = .56, SE = .06, β = .52, $p < .001$) but not with EE (B = .10, SE = .07, β = .07, $p = .187$). The regression model for **loneliness** was significant, $F(2,463) = 4.88$, $p = .008$, adj. $R^2 = .016$, with significant model contributions from EE (B = -.25, SE = .10, β = -.17, $p = .010$) but not CC (B = .04, SE = .07, β = .04, $p = .552$). Finally, **perceived companionship provision** was positively predicted, $F(2,464) = 186.86$, $p < .001$, adj. $R^2 = .444$. Both factors positively contributed: EE (B = .46, SE = .07, β = .33, $p < .001$) and CC (B = .42, SE = .05, β = -.39, $p = .001$). Amount of **time spent weekly** interacting with the companion was also predictive for both factors: $F(2,445) = 20.85$, $p < .001$, adj. $R^2 = .082$. Both factors positively contributed: EE (B = 1.105, SE = .39, β = .18, $p = .005$) and CC (B = .64, SE = .29, β = .14, $p = .028$). How long it had been since the AIC was created (i.e., its age in months), however, was not associated, $F(2,464) = 2.09$, $p = .125$, adj. $R^2 = .005$.

Altogether, these results are taken as evidence that the scale generally exhibits construct and predictive validity, with noted deviations addressed in *post-hoc* analysis and discussion.

**Table 2**
Machine Companionship Factor Scores, Compared Across Relationship Labels

|    | Intimate ($n = 193$) | Advisor ($n = 144$) | Associate ($n = 40$) | Acquaintance ($n = 56$) |
|----|----------------------|---------------------|----------------------|-------------------------|
| EE | 5.62 (1.19)c        | 5.43 (1.06)b,c      | 4.96 (1.19)a,b       | 4.65 (1.21)a            |
| CC | 4.90 (1.54)c        | 4.11 (1.58)b        | 3.18 (1.30)a         | 3.83 (1.62)a,b          |

*Note:* Values presented are *M* (*SD*). Subscripts represent membership in homogenous subsets, role types with the same subscript are not statistically different in scores for that row.

### 4.4 *Post-Hoc* Explorations

To facilitate future engagement of the AIC scale, additional relationships are explored across the gathered descriptive variables.[4] Participant **age** significantly predicted companionship factors, $F(2,464) = 5.91$, $p = .003$, adj. $R^2 = .021$, with EE positively associated with age in years (B = 2.18, SE = .63, β = .22, $p < .001$) and CC negatively associated with age (B = -1.13, SE = .47, β = -.15, $p = .017$). **Number of companions** was not predictive of EE/CC $F(2,463) = 2.97$, $p = .052$, adj. $R^2 = .008$. Considering the **AI platform type**, ANOVA of EE and CC across platform indicates a significant difference, Pillai's trace = .125, $F(4,928) = 15.48$, $p < .001$. *Post-hoc* Scheffé contrasts for EE indicate only one pair were significantly—and only slightly—different with LLM scores being highest ($M = 5.44$, $SD = 1.10$) and companion apps lowest ($M = 4.93$, $SD = 1.51$) respectively. Contrasts indicate CC scores did not differ across platform types. Finally, because one or the other of the two factors sometimes exhibited unexpected (non)relations for construct-validation variables, it is possible that although EE and CC are correlated, different patterns in values could signal different experiential outcomes. To explore this possibility, I conducted a canonical correlation analysis (see Sherry & Henson, 2005)—one that identifies correlations between *sets* of variables. between the variable sets) and observe an overall very large effect of $R_c^2 =$ The overall model was significant, $F(18, 712) = 59.69$, $p < .001$, Wilks's λ = .159, such that I reject the null (that there is no relationship between the variable sets) and observe an overall very large effect of $R_c^2 = .84$. (This is interpreted as one would consider a multiple $R^2$ in a regression, as the "proportion of variance shared across all variable sets across all functions"; Sherry & Henson, 2005, p. 42). Examination of the canonical correlations indicates two functions, both of which are significant at $p < .001$. I elect to interpret both as they both explain a large amount of variance between their orthogonal functions (76.15% and 33.38% respectively, noting that function 2 is created after function 1 and so explains the *remaining* variance).

---
[4] Current U.S. law limits data collection and reporting to binary sex for federally funded studies. I have elected to rely on the platform's binary-sex sampling mechanism only to facilitate diversity in the sample; other comparisons are not made here and I acknowledge this as a limitation of the study.

In examining the individual sets, the rule of thumb is to interpret only structure coefficients (function weights, or $r_s$) above $|.30|$ as contributing substantively to the function (i.e., 9% of variance). However, the goal of this analysis is to explore relationships between the MC scale factors and the validation set and the lowest function loading for MC is the function 2 EE loading of -.232, $|.232|$ is set as the minimum benchmark for interpretation. For the criterion set (EE and CC dimensions of AI companionship): In function 1, both EE and CC contribute positively and strongly to the function. In function 2, EE contributes negatively while CC contributes positively and more weakly. In other words, function 1 represents strong EE/CC relations while function 2 represents a moderate CC and diminished EE. For the predictor set (validation variables), in function 1, nearly all variables meet the benchmark and most are moderate to strong contributors to that latent variable. For function 2, experiential mind, IOS, and relatedness need satisfaction are positive contributors alongside a negative contribution from utilitarian motivations. See Table 3.

Altogether, there two distinct patterns in latent variable correlations among participants' AIC relations, marked by orthogonal MC functions. Function 1 represents a form of MC in which EE/CC are balanced and are, together, correspondent with both social and utilitarian experiences. Function 2 represents MC marked by reduced EE and increased CC, corresponding with perceptions of the AI's experiential mind, IOS, and the satisfaction of relational needs and a reduction of utilitarian motivations. I interpret this to suggest there may be two experiential templates for MC. One appears to be *socioinstrumental* in which the relation is based on both self-elevating exchanges and coordination, serving both relational and functional needs but is ultimately driven by utilitarian motives; the machine is a useful, mindful agent is synergistic with and advancing one's interests. The other is *autotelic* in which the relation is marked by coordination as a form being-together and the machine is seen as an experiencing entity who satisfies primarily relatedness needs, and the human rejects instrumental motivations and self-elevating experiences. These potentials are considered further in the discussion.

### 4.5 Confirmatory Analyses

Following best practices in scale development, the two-factor solution was tested with a separate sample of AI users ($N = 249$, benchmarking at >20 participants per scale item). Data collection used the same sampling parameters, excluding participants from the first data collection. Measures and procedures were identical, except for the MC scale: Items composing the EE and CC factors were retained and the others from the original pool were removed. Measures were similarly reliable with each scale at $\omega > .80$ (see online supplements).

Using AMOS 30 and maximum likelihood estimation, confirmatory factor analysis tested the two-factor model (Figure 1) according to fit indices as recommended by Bowman and Goodboy (2020). Items loaded onto the two factors as expected, and factor loadings were statistically significant ($p < .01$) with standardized weights ranging from .682 to .864. Global fit indices were mixed: $\chi^2(53) = 179.97$, $p < .001$ (the hypothesized model is distinct from a saturated model, acknowledging $p$ is sensitive to sample size), SRMR = .051 (good fit), RMSEA = .098, 90% CI [.083, .114], $p < .001$ (adequate fit, but not ideal), and CFI = .942 (good fit, with ~94% improvement over the baseline model). Because of the mixed results, standardized residuals were examined to identify local-fit issues. Two borderline item pairs

**Table 3**
Canonical Correlations Between MC Factor-sets and Validation Variable-sets

| | Function 1 | | | Function 2 | | | |
|---|---|---|---|---|---|---|---|
| Variable | Coef | $r_s$ | $r_s^2$ (%) | Coef | $r_s$ | $r_s^2$ (%) | $h^2$ (%) |
| EE | .743 | **.973**✓ | 94.7 | -1.195 | **-.232**✓ | 05.3 | 100.0 |
| CC | .327 | **.849**✓ | 72.1 | 1.368 | **.528**✓ | 27.9 | 100.0 |
| $R_c^2 = .84$ | | | | | | | |
| Utilitarian M. | .399 | **.793**✓ | 62.9 | -.654 | **-.478**✓ | 22.8 | 85.7 |
| Mind – Exper. | .078 | **.553** | 30.6 | .562 | **.593**✓ | 35.2 | 65.8 |
| Mind – Agency | .210 | **.790** | 62.4 | -.307 | .003 | 00.0 | 62.4 |
| IOS | .200 | **.779** | 60.7 | .438 | **.350**✓ | 12.3 | 73.0 |
| Positive Affect | .105 | **.592** | 35.1 | -.165 | **-.132**✓ | 01.7 | 36.8 |
| Relatedness N.S. | .125 | **.603** | 36.4 | .411 | **.414**✓ | 17.2 | 53.6 |
| Loneliness | -.047 | -.134✓ | 01.8 | .089 | .109 | 01.2 | 3.0 |
| Companionship | .211 | **.753**✓ | 56.7 | .015 | .191✓ | 03.7 | 60.4 |
| Weekly Hours | .051 | **.334** | 11.2 | -.122 | .023✓ | 00.0 | 11.2 |

*Note:* EE = Eudaimonic Exchange, CC = Connective Coordination. Coef = standardized canonical function coefficients (function weights). $r_s$ = structure coefficients (correlation between the variable and the latent variable in the other set). $r_s^2$ = squared structure coefficients (% shared variance between that variable and the latent variable in the other set). $h^2$ = communality coefficients (amount of variance in variable summed across functions). Structure coefficients above the interpretation threshold are indicated in bold. ✓ indicates the general pattern (of meeting or failing the interpretation benchmark *and* matching valence) was replicated in the subsequent validation sample.

were observed, however they were both retained because residuals were below the standard |2.58| threshold and the items were semantically and conceptually distinct.

Finally, because of the high correlation between factors ($r = .77$, similar to the first sample), the two-factor model was tested against an alternative one-factor model. The chi-square difference test indicated that the two-factor model provided significantly better fit, $\Delta\chi^2(\Delta df = 1) = 257.43$, $p < .001$ (one-factor RMSEA = 1.015, indicating poor fit compared). Altogether, analyses indicate the data was a *reasonable* fit to the model and the two-factor model is superior to an alternative one-factor model.

**Figure 1**
Hypothesized Two-factor Model for MC, with Factor Loadings

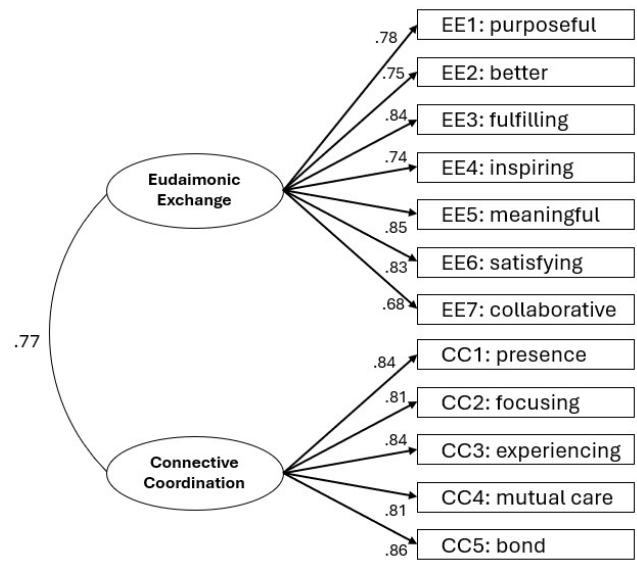

*Note:* EE = Eudaimonic Exchange, CC = Connective Coordination

With the retained model, construct validity tests were again conducted. Patterns observed in the first sample were generally replicated, with some exceptions. Whereas the prior data demonstrated a positive link between agentic mind perception and both factors, CC was uncorrelated here. Relatedness need satisfaction was previously only positively correlated with CC, in the second sample it correlated with both factors. Whereas loneliness was previously negatively correspondent with EE (and not related to CC), here there was no link with either factor. Finally, where platform-type differences were previously observed (EE lowest for AIC apps and highest for LLMs), no differences were observed in the second sample. See online supplements for complete details.

Performing a canonical correlation of MC factors compared to validation variables, the model was again significant, $F(18, 382) = 32.76$, $p < .001$, Wilks's $\lambda = .155$, with two significant functions and a very large effect of $R_c^2 = .85$. For this dataset, function one similarly had positive contributions from both MC factors, however the function weights were lower (EE $r_s = .708$ versus .973; CC $r_s = .367$ versus .849), while the function weights for function two were much higher (EE $r_s = -1.207$ versus -.232; CC $r_s = 1.350$ versus .528). Among construct validation variables, the original patterns were partially replicated. Function one similarly had positive contributions from utilitarian motivations and heuristic companionship perceptions, however the other prior positive contributions did not meet the interpretation threshold. Function 2 replicated most contributions, adding a positive contribution of agentic mind perception. See Table 3 for notation of these patterns. As this data *generally* follow the same pattern as in the first data set, these values are interpreted as reasonably replicating the MC templates. Across both samples, data indicate the existence of:

- A socioinstrumental MC template:
  ($EE^+$ and $CC^+$) *correlate with*
  (Utilitarianism$^+$ and Companionship$^+$)
- An autotelic MC template:
  ($EE^-$ and $CC^+$) *correlate with*
  (Utilitarianism$^-$ and Experience mind$^+$ and IOS$^+$ and relatedness need satisfaction$^+$)

Further, these templates are reliably *not* linked to loneliness. However, deviations from patterns in the first sample indicate potential instability in how mind perception, self-other overlap, and relatedness need satisfaction may vary in relation to MC factors.

## 5 Discussion

This study inductively devised and initially validated a self-report scale to measure machine companionship experiences with varied conversational AI. The induction was grounded in a five-facet definition of MC and produced a scale solution with two factors—Eudaimonic Exchange and Connective Coordination—and its construct validity was generally supported by expected correspondence with mind perception, self-other overlap, positive affect, relatedness need satisfaction, heuristic companionship perception, utilitarian motives (negative) and loneliness (negative or nonsignificant). EE and CC capture *perceived active operations* of MC that are distinct from companionships' antecedents and outcomes. Exploration of unexpected patterns in validation analyses, findings incidentally suggest two templates for AIC relations.

### 5.1 Characterizing MC Scale Dimensions

The scale solution comprised two dimensions that require interpretation. The first, named **Eudaimonic Exchange (EE)**, included items reflecting positive affect related to psychological elevation (fulfilling, inspiring, meaningful, satisfying) as well as purposiveness in interaction (purposeful exchange, collaborative). Together, these items exemplify an intersection of two key notions from psychological studies of well-being. The first is *eudaimonic well-being*, where eudaimonia is a term originating in Aristotelian philosophy referring to the promise of humans to achieve excellence in line with their own unique potential, culminating in a state of optimal self-realization (Aristotle, 350 BCE/2009). The notion has since been engaged to conceptualize psychological well-being as positive human functioning (see Ryff & Singer, 2008) as this elevation manifests purpose, growth, autonomy, mastery, positive relations, self-acceptance (Ryff, 1989). The second anchoring line of research attends to *growth-fostering relationships*, that supports moving eudaimonic elevation beyond an exclusively solitary affair to one potentially animated by our interactions with others. Growth-fostering relationships are those marked by mutual empathy and authenticity, and empowerment and well-being improve as a result, including enhanced self-worth, vitality for life, and clarity in identity (Miller, 1986).

The second factor, **Connective Coordination (CC)**, included items reflecting dyadism (mutuality, bonding) and co-activity (co-presence, mutual attention, shared experience). These items cohere around the notion of interpersonal synchronization, or the extent to which dynamic systems (here, machine and human) have correspondent actions, thoughts, and feelings over time as they are coupled to manifest a more complex system (the relationship) that has distinct qualities (Vallacher & Nowak, 2017). In other words, relationship partners have correspondent internal and external states. Indicated by included scale items, these states range from mere being-with as the subjective experience of shared social context (Campos-Castillo & Hitlin, 2013) to states suggesting a perceived shared reality through conspicuous co-behavior (see Rossignac-Milon & Higgins, 2018; see also Liang & Banks, 2025). CC may be an intuited experience of what Harvey and Omarzu called the "minding" of relationships (1997, p. 225), or the engagement of behaviors that facilitate generating knowledge of the other and the engagement of that knowledge to further the relationship—behaviors like verbal and nonverbal expressions that can animate a rapid expansion of self-concept to include the other (Aron & Aron, 1996). That minding behaviors often take the form of told stories and self-disclosures illuminates how this form of experienced intimacy may unfold with AICs, since textual or verbal exchanges are currently the primary forms of communication in MC. Companion technologies both elicit and afford self-disclosure and permit each to "read" the other, creating the conditions for seeing the other as a legitimate actor and for making inferences about the causes of their behaviors (Harvey & Omarzu, 1997, p. 226)—and perhaps for the perception of other minding operations of acceptance, reciprocity, and continuity.

## 5.2 Socioinstrumentality and Autotelicity as Relational Templates

There were unexpected patterns in initial analyses: EE is related to utilitarian motives, has overlapping associations with intimate/advisor/associate AI roles, was highest for those considering general-purpose LLMs, and had no correspondence with perceived experiential mind or relatedness-need satisfaction. This suggests the eudaimonics of EE are not necessarily experienced as the self-transcendent variety (Haidt & Morris, 2009), but instead as personal fulfillment and accomplishment accomplished through or with the relationship (see Ryan & Martela, 2016)—and consciously adopted to that end. Additionally, CC had no correlation with positive affect. These outcomes prompted exploratory analysis, treating the MC factors and validation variables as candidates for latent variable sets, and outcomes signal there may be two possible relationship archetypes as the two MC factors combined differently to signal distinct relational templates. Those templates were reasonably replicated in a separate sample.

The first template is one for socioinstrumental companionship in which EE and CC positively correspond with indicators of seeing the AI as providing companionship *and* highly correlated with utilitarian motivations (and sometimes mindful). This supports the suggestion that a form of MC may abide based on self-interested rather than self-transcendent eudaimonics—in other words the companionship may be more transactional and produce more functional benefits (e.g., Bayor et al., 2025), so the AI is agentic and useful. The second is an autotelic companionship in which CC and EE are inversely related, with CC (positive contributor) corresponding with seeing the AI having mindful experience (though not necessarily agentic), relatable, self-relevant agent engaged as intrinsically valuable (i.e., a rejection of utilitarian motives). This template indicates a form of MC more aligned with Banks and Li's (2025) conceptualization emphasizing autotelicity, where companionship is a valid thing-in-itself. There is a caveat, though: The absence of the AI's perceived mindful agency suggests (in line with dyadic perspectives on morality; Gray & Wegner, 2012) that the experiencer/non-agent AI is a *patient* to the human's actions and desires. In other words, the AI may be seen as having legitimate experiences of the world and as contributing to self and relatedness, but as a non-agent it does so as a *subject* of the human's engagement.

The emergence of both socioinstrumental and autotelic templates, in some respects, is not surprising, given that many of our everyday human-human relations are both social and functional (or ecological and egoistic; Crocker & Canavello, 2017). Other works have identified for MCs both self-improvement and autotelic use motivations, emotional and productive conversation topics (Liu et al., 2025), and that people take up AICs for interaction but also see them as tools (Common Sense Media, 2025). These orthogonal templates may be able to inform the debate about the presence or absence of mutuality and reciprocity in these relationships.

## 5.3 Limitations and Future Directions

As with any scale development effort, this study carries inherent limitations—focus on a single type of companionable machine, limiting to English-speaking participants, mild inconsistency between samples—and requires further validation. In particular, it is well-understood that experiences and understandings of *both* social machines and close relationships are shaped by cultural norms and practices (see Kou & Zhang, 2024) such that cross-cultural and translational validation is required for the scale to be valid for non-U.S./U.K. populations. Additionally, there was a somewhat unexpected prevalence of LLM users in both samples, which may have animated patterns distinct from those that might emerge for other interaction platforms. Although this solution maintained factor coherence across the three platform types, future work must explore how observed factors and templates may vary across varied forms of companionable machines like social robots and voice assistants.

These limitations acknowledged, the derived machine companionship scale is grounded in past systematic reviews, demonstrates general construct validity properties, and was incidentally useful in identifying MC templates—indicating it will be useful for a host of future investigations. Perhaps most important for advancing empirical understandings of MC, this work distinguishes the *subjective relational state* of MC from its antecedents (e.g., motivations, preferences, loneliness) and effects (e.g., need satisfaction, wellbeing outcomes) so we can more carefully examine their causal relations. For instance, considering the aforementioned debate over MC and loneliness, neither factor and neither template indicates correspondence between MC and loneliness; this data meet other recent findings in suggesting that companionship (variably measured) does *not* directly correspond with loneliness (e.g., Liu et al., 2025) or is negatively related (e.g., Ebner & Szczuka, 2025) as an antecedent or outcome.

Among myriad potential applications of this scale, some scholars suggest we explore the mechanisms by which people strategically engage AICs as "quasi-domesticated" objects in service to and integration with human lives (Neff & Nagy, 2025). Others consider whether people opt for the comfort of an always-available and low-conflict AIC, not out of escaping human companionship but instead by inviting AI to fill relational roles "that humans have vacated or bungled" (Safronov & Elio, 2025). Some evidence indicates MC experiences prompt deep existential questions and hopes for the productive AI relationships (Banks et al., 2025), and other findings point to concern over harmful AI behaviors ranging from boundary violations to abuse (Zhang et al., 2025). These and other questions about anticipated discord and harmony among humans and machines may be productively explored through the lens of perceived Eudaimonic Exchange and Connective Coordination. Future scholarship should explore these potentials—and the identified templates—as important relational experiences.

Finally, I draw on the observed templates to challenge whether findings call for reconsideration of the initial conceptual approach MC. I drew from Banks & Li's (2025) definition: "an autotelic, coordinated connection between a human and machine that unfolds over time and is subjectively positive" (p. 13). Although coordination, dyadic connection, and positivity were supported (and temporality is inherent), the *requirement* of autotelicity was partially challenged here. One relational template was decidedly grounded in that notion, seeing cohering emphases on CC's form of being-with an experiencing entity. However, the socioinstrumental template was *highly correspondent* with utilitarian motivations alongside the other expected social-cognitive and relational variables. At minimum, this finding requires definitional acknowledgement that the connection may be autotelic *and/or self-interested*, though the exact nature of that interestedness requires further exploration. Beyond this, the field requires a theoretical infrastructure to simultaneously and synergistically consider the authentically social experience and the technologically functional experience inherent to companionable machines—one where (as Seibt et al. [2020] suggest) it may not be prudent to hold human companionship as the standard.

## AUTHOR NOTE

Thanks to: Dr. Caleb T. Carr for his collaboration in an early version of this work; Drs. James Cummings, Jessica Szczuka, Nicholas David Bowman, and Laura Kunold for their formative feedback; Zhixin Li for support in the data collection process. This work is funded by NSF-CISE under grant 2401591. All sampling, coding, and analysis documentation is available in the online supplements for this project: https://osf.io/hjg2a.